\documentstyle[prb,aps,multicol,12pt,epsf]{revtex}

\author{S. N. Dorogovtsev\cite{E}}
\title{Avalanche mixing of granular solids in a rotating 2D drum \\
and discrete mapping
}
\draft
\address{A.F.Ioffe Physico-Technical Institute, 194021 St.Petersburg, Russia
}

\begin{document}

\maketitle
\begin{abstract}
Evolution of mixing of granular solids in a slowly rotated 2D drum is
considered as a discrete mapping.  The rotation is around the axis of
the upright drum which is filled partially, and the mixing occurs only
at a free surface of a material. The most simple cases that demonstrate
clearly the essence of such a type of mixing are studied analytically.
We calculate the characteristic time of the mixing and the distribution
of the mixed material over the drum.
\end{abstract}

\pacs{PACS numbers: 64.75, 46.10}


By the term "avalanche mixing" \cite{metcal} one calls the mixing
of granular solids that proceeds {\it only} at a free surface layer.
Mixing occurs in a slowly rotating upright drum (we shall consider
only a two dimensional drum) which is filled partially
(see fig.~1).
Granules mix in avalanches falling from the upper half of a free surface
to the lower half. In spite of the simplicity of the process,
nature of the avalanche mixing is nontrivial and experimental mixing
patterns look striking \cite{metcal}. Nevertheless, nature of the
avalanche mixing can be understood using purely geometrical approach
\cite{metcal,dor1,peratt1,dor2,peratt2}. Note that though mixing and
segregation in rotated drums is studied intensively
\cite{buch,po,cant,koh,ris1,ris2,ris3,raj,baum2,baum,baum3,zik,hil,clem,can2}
, the avalanche mixing is of special interest, because of clearness
of the problem formulation.

If one assumes that the difference
between the angles of repose and marginal stability is little
(the case of the finite difference is studied thoroughly by
B. A. Peratt and J. A. Yorke \cite{peratt1,peratt2}) it is
possible to describe the mixing analytically \cite{dor1,dor2}.
(Note that small granules of the different fractions are distinguished only
by color.)
The fastest possible avalanche
mixing takes place if fractions mix completely in avalanches, so the
granular material is always mixed at the lower half of a free surface.
The experiment \cite{metcal} appears to be surprisingly close to this
extremal regime \cite{dor2}. Below we shall consider an opposite extremal
situation in which granules, while flowing along a free surface
are not mixed at all. One may ask: is there any mixing in such a situation
at all? We shall give a positive answer. The mixing exists
(apart from a center part of the drum, of course), and it is
rather efficient.
Indeed, the fractions will be finally mixed!
We shall describe dynamics of such a mixing.
\begin{figure}[\!h]
\epsfxsize=3.5in
\epsffile[-128 233 257 569]{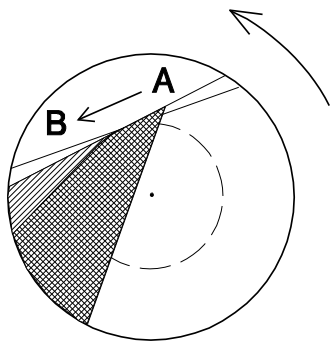 }
\caption{
\narrowtext
The avalanche mixing scheme.
For an infinitely small turn of the drum, the granules of different
fractions flow from sector $A$ to sector $B$, undergoing mixing.
The free surface of the material is tilted at the angle of repose all the
time. The pure fractions are shown by the black and white colours.
Regions of mixed material are denoted by the grey colour.
The degree of mixing is not shown. The material inside of the dashed
circle will be never mixed. The result of the avalanche mixing depends on
the way in which the granules flow from sector $A$ to sector $B$.
\label{fig1}}
\end{figure}

\begin{figure}[\!h]
\epsfxsize=3.5in
\epsffile[28 133 557 709]{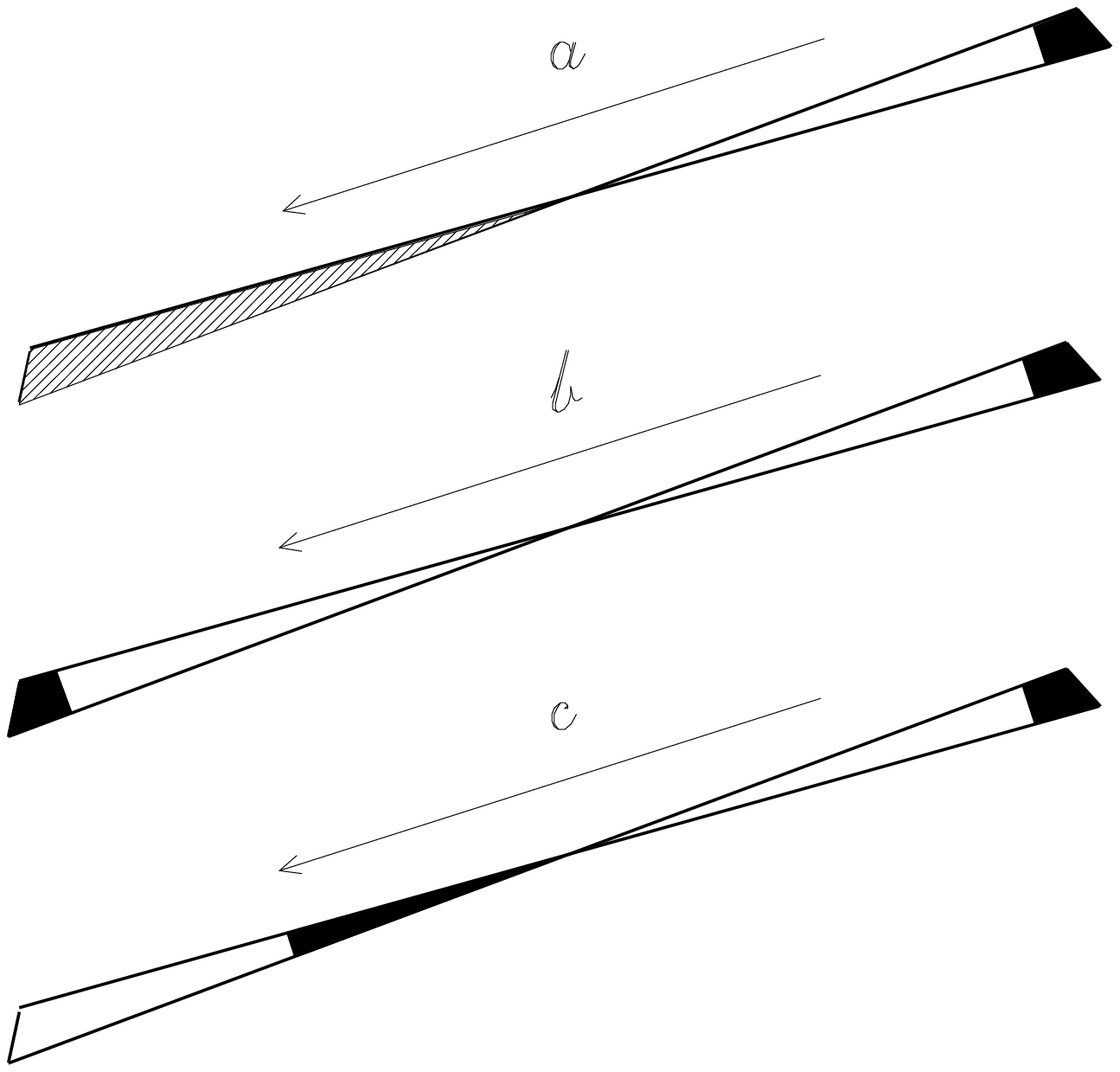 }
\caption{
\narrowtext
Limit regimes of the evolution of granules while flowing from
the upper half of a free surface to the lower half.
\\
$a$ -- the material mixes completely in the avalanches;
\\
$b$ -- after the flow, the material appears to be inverted, and
upper granules become lower ones -- the $x^\prime=x$ transformation
of the wedges
($x$ is a linear coordinate at a free surface, counted from the midpoint
of the free surface in both directions).
\\
$c$ -- the material slips as a whole from the upper half of a free surface
to the lower half, so lower granules stay to be lower ones after the slip.
Using the condition of volume conservation one obtains easily the
$x^\prime=\protect \sqrt{\sin^2\theta-x^2}$ transformation of the wedges.
Here the angle $\theta$
characterizes the relative volume of an empty space in the drum.
\label{fig2}}
\end{figure}

Then there are two possible cases. In the first one, granules with the
linear coordinate $x$ at the upper half of a free surface (we assume that
at the free surface midpoint $x=0$) fall to the point $x^\prime=x$ at the
lower half, so in fact the surface wedge is not changed after avalanches
(see Fig.~\ref{fig2},$b$).

In the second case, the material from the upper wedge slips to the lower
one as a whole (Fig.~\ref{fig2},$c$).
The drum radius is unit, and the measures of the amount
of the material are the angles $\theta \leq \pi/2$ for more then half
filled drum and $\vartheta=\pi-\theta \leq \pi/2$ for less then
half filling (see Fig.~\ref{fig1}).
Thus, the coordinates of the lower and the upper points
of a free surface  will be $\sin\theta$ or $\sin\vartheta$.
Then, accounting for volume conservation, one describes
the transformation of the coordinates as $x^\prime=\sqrt{\sin^2\theta-x^2}$
(or $x^\prime=\sqrt{\sin^2\vartheta-x^2}$) in these situations.

Introducing the scaled coordinate $z \equiv x/\sin\theta$ or
$z \equiv x/\sin\vartheta$ we get for the cases under consideration:
$z^\prime=z$ and $z^\prime=\sqrt{1-z^2}$.

Let us describe by the number $k$ events of appearing of some granule
at the lower half of a free surface. The corresponding coordinate
of the granule
at the lower part of a free surface and time will be $(x_k,t_k)$.
(The angle of the drum turn plays the role of the time.)
Transformation of this pair describes completely the mixing, since
all the granules move as a whole in the bulk of the material.
We shall study this discrete mapping.

We start with consideration of the $z^\prime=z$ transformation
case for less then half
filled drum. One can see from Fig.~\ref{fig1}
that the coordinate is unchanged
during the mapping and $t_{k+1}=t_k+2\arctan(x_k/\cos\vartheta)$.
\begin{figure}[\!h]
\epsfxsize=3.5in
\epsffile[28 133 557 509]{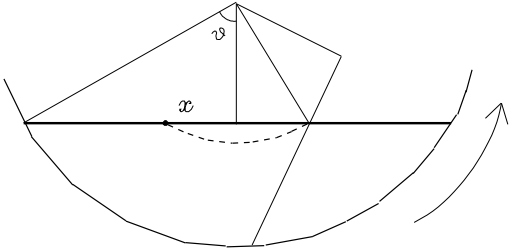}
\caption{
\narrowtext
The avalanche mixing in the event of a less then half filled drum
and the $x^\prime=x$ transformation of the wedges.
A free surface is shown to be horizontal, since our
results do not depend on the angle of repose.
The angle $\vartheta \leq \pi/2$
characterizes the relative volume of the granular material.
If a granule is at the point $x$ at the left half of a free surface,
it will always stay at a dashed circle sector.
\label{fig3}}
\end{figure}
Thus, one obtains the following trivial discrete mapping \cite{meiss}:

\begin{equation}
\label{M1}
(z_{k+1},\,t_{k+1})=(z_k,\,t_k+2\arctan(z_k \tan\vartheta)) \ .
\end{equation}

\begin{figure}[\!h]
\epsfxsize=3.5in
\epsffile{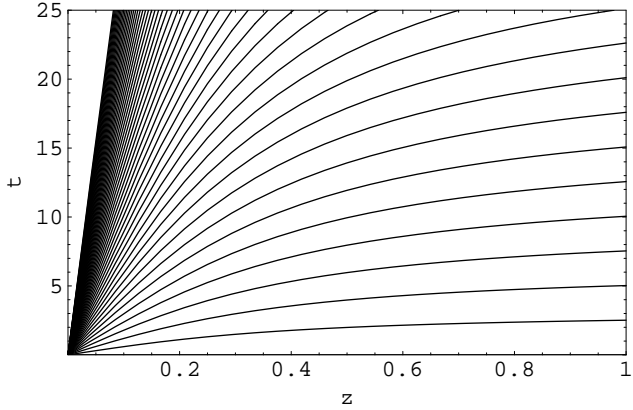 }
\caption{
\narrowtext
Evolution of a thin layer of the black fraction for a less
than half filled drum in the case of the $z^\prime=z$ transformation
$(z\equiv/x\sin\vartheta)$. $\vartheta=0.4\pi$. (See Eq.~(\protect\ref{M2})).
Crossings of the line $t=const$ with the curves correspond to
the points at the lower half of a free surface in which black granules
appears at the time $t$.
\label{fig4}}
\end{figure}

One may study the simplest case of an infinitely thin layer
of the black fraction over the white fraction at the initial moment,
that is $(z_0,t_0)=(z,0)$. Then the black material will evolve in
the following manner:

\begin{equation}
\label{M2}
(z_k,\,t_k)=(z,\,2k\arctan(z \tan\vartheta)) \ .
\end{equation}
Thus, the black granules will be at the point $z$ of a free surface
at moments $t_k(z),\ k=0,1,2\ldots$
(see Fig.~\ref{fig4}).

Formally speaking, there will be no true mixing of the material at
any time, but near the fixed $z$ the distance between black granule
layers are diminished with time. When, in the vicinity of $z$,
this distance becomes to be
of the order of the granule diameter, the material is
in fact mixed in this area. A very simple geometrical consideration
gives for this distance:

\begin{equation}
\label{M3}
d=\frac2 t \frac{z\sin\vartheta
(1+z^2\tan\vartheta)\arctan^2 (z\tan\vartheta)}
{\{\left[1+(1+z^2\tan^2\vartheta)\arctan(z\tan\vartheta)/t  \right]^2+
z^2\tan^2\vartheta\}^{1/2}} \ .
\end{equation}
If now $d$ is the relative diameter of granules (recall that the drum radius
equals $1$)
then one obtains the corresponding characteristic time

\begin{equation}
\label{M4}
t_{mix}=\arctan(z\tan\vartheta)
\left\{\left[\frac{4z^2\sin^2\vartheta}{d^2}
(1+z^2\tan\vartheta)\arctan^2(z\tan\vartheta)-
z^2\tan^2\vartheta   \right]^{1/2}
-1 \right\} .
\end{equation}
The expression is most simple in the following limit cases.
If $z\tan\vartheta \ll 1$,

\begin{equation}
\label{M5}
t_{mix}=\frac{2 z^3}{d} \tan^2\vartheta \sin\vartheta \ .
\end{equation}
If $z\tan\vartheta \gg 1$,

\begin{equation}
\label{M6}
t_{mix}=\frac{\pi^2 z^2}{2d} \tan\vartheta \sin\vartheta \ .
\end{equation}

Finally, the black fraction from the point $z$ will be distributed
homogeneously over the dotted line shown in Fig.~\ref{fig3}. If initially
the black fraction occupies the sector of the $\gamma\to 0$ angle
at the lower half of a free surface, then the concentration of
the black fraction in the mixture will be

\begin{equation}
\label{M7}
c(z)=\frac{\gamma}{2\arctan(z\tan\vartheta)} \ ,
\end{equation}
so the mixing differs greatly for different $z$.
As it can be seen from Eq.~(\ref{M6}), when $\vartheta \to \pi/2$,
$t_{mix}\to\infty$, and any mixing is practically absent
(see Fig.~\ref{fig4}).
\begin{figure}[\!h]
\epsfxsize=3.5in
\epsffile{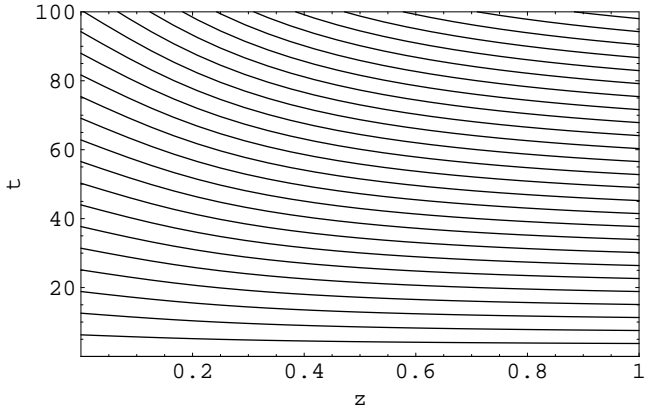 }
\caption{
\narrowtext
Same as Fig.~\protect\ref{fig4}
for a more than half filled drum in the case
of the $z^\prime=z$ transformation $(z\equiv\sin\theta)$. $\theta=0.4\pi$.
(See Eq.~(\protect\ref{M8})).
\label{fig5}}
\end{figure}

The case of a more then half filled drum
$\theta \equiv \pi/2-\vartheta \leq \pi/2$
can be considered in the same manner. In the event $z^\prime=z$ one
writes

\begin{equation}
\label{M8}
t_k=2k[\pi-\arctan(z \tan\theta)]
\end{equation}
(see Fig.~\ref{fig5}) -- cf Eq.~(\ref{M2}).
The material in the point $z$ will be mixed to a "homogeneous" state
with the black fraction consentration

\begin{equation}
\label{M9}
c(z)=\frac{\gamma}{2\pi-2\arctan(z\tan\theta)} \ ,
\end{equation}
in a time

\begin{equation}
\label{M10}
t_{mix}=[\pi-\arctan(z\tan\theta)]
\left\{ \left[\frac{4z^2\sin^2\theta}{d^2}(1+z^2\tan^2\theta)
\left[\pi-\arctan(z\tan\theta)\right]^2
-z^2\tan^2\theta \right]^{1/2}-1 \right\} .
\end{equation}
Thus, if $z\tan\theta \gg 1$,

\begin{equation}
\label{M11}
t_{mix}= \frac{\pi^2}{2d}\tan\theta\sin\theta
\end{equation}
and in the event of  $z\tan\theta \ll 1$

\begin{equation}
\label{M12}
t_{mix}=\frac{2 \pi^2 z}{d}\sin\theta \ .
\end{equation}
Note, that Eqs~(\ref{M10}) and (\ref{M12}) are valid only if
there are many crossings
of the black fraction trace with the lower half of a free surface, i.e.
if $t_{mix} \gg 2\pi(2\pi/2\theta)$. Thus, as it follows from
Eq.~(\ref{M12}), the $\theta$ angle should be
$\theta \gg\sqrt{d/z}$. Otherwise $t_{mix} \sim 2\pi^2/\theta$.

\begin{figure}[\!h]
\epsfxsize=3.5in
\epsffile{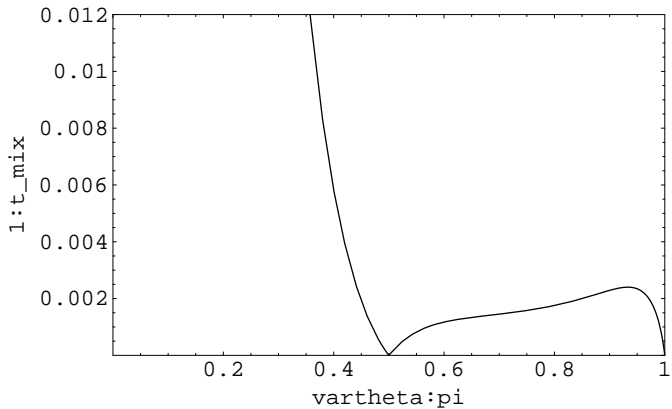 }
\caption{
\narrowtext
$\protect t_{mix}^{-1}$ vs $\vartheta/\pi$.
The later one characterizes the
relative volume of the granular material: the drum is empty, half
filled, or full when $\vartheta=0,\pi/2,$ or $\pi$, correspondingly.
The crossover region near $\vartheta\sim 1$ is shown by guess-work.
$z=0.5,d=0.01$.
\label{fig6}}
\end{figure}

We see that there is a minimum of $t_{mix}$ at some intermediate filling
value. A full dependence of $t_{mix}$ on $\vartheta$ (or rather
$t_{mix}^{-1}(\vartheta)$) is shown in Fig.~\ref{fig6}.
It somewhat resembles the characteristic mixing time behavior
in the case of a full mixing in the wedges \cite{metcal}, though
the meaning of $t_{mix}$ is quite different now.

\begin{figure}[\!h]
\epsfxsize=3.5in
\epsffile{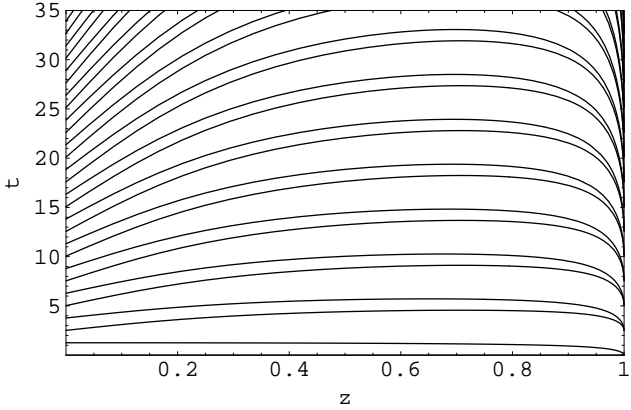 }
\caption{
\narrowtext
Same as Figs~\protect\ref{fig4} and \protect\ref{fig5}
for a less than half filled drum and
the $z^\prime=\protect\sqrt{1-z^2}$ transformation of the wedges.
$\theta=0.4\pi$. (See Eq.~(\protect\ref{M15})).
\label{fig7}}
\end{figure}

Now we proceed with the case of a less then half filled drum and the
$z^\prime=\sqrt{1-z^2}$ transformation. One see that

\begin{equation}
\label{M13}
(z_{k+1},\,t_{k+1})=
\left(\sqrt{1-z_k^2},\,t_k+2\arctan(z_k \tan\vartheta)\right) \ ,
\end{equation}
so if initially $(z_0,t_0)=(z,0)$, then

\begin{eqnarray}
\label{M14}
(z_{2n},\,t_{2n})=
\left(z,\,2n\arctan(z_k \tan\vartheta)+
2n\arctan(\sqrt{1-z_k^2} \tan\vartheta)\right)
\phantom{eeeeeeeeeeee}
\nonumber\\
(z_{2n+1},\,t_{2n+1})=
\left(\sqrt{1-z^2},\,2(n+1)\arctan(z_k \tan\vartheta)+
2n\arctan(\sqrt{1-z_k^2} \tan\vartheta)\right) \ ,
\end{eqnarray}
where $n=0,1,2,\ldots$. Therefore,

\begin{eqnarray}
\label{M15}
t_{2n}=2n\arctan(z\tan\vartheta)+2n\arctan(\sqrt{1-z^2}\tan\vartheta)
\phantom{eeeeeee}\nonumber\\
t_{2n+1}=
2n\arctan(z\tan\vartheta)+(2n+1)\arctan(\sqrt{1-z^2}\tan\vartheta)
\end{eqnarray}
(see Fig.~\ref{fig7}).

Now there is a point $z=1/\sqrt{2}$ in which the fractions
will be never mixed, and mixing in fact occurs in the vicinity of
$z=0$ or $z=1$. The region of poor mixing is broaden as the drum filling
tends to the one half level. The averaged concentration of the black
fraction is

\begin{equation}
\label{M16}
c(z)=
\frac{\gamma}{\arctan(z\tan\vartheta)+\arctan(\sqrt{1-z^2}\tan\vartheta)}
\end{equation}
(cf Eq.~(\ref{M7})).

\begin{figure}[\!h]
\epsfxsize=3.5in
\epsffile{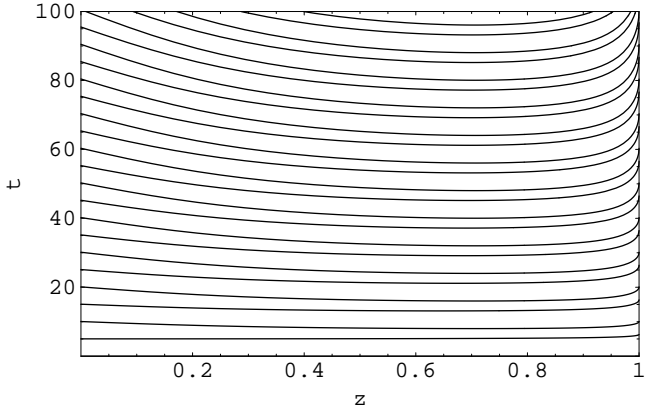 }
\caption{
\narrowtext
Same as Figs~\protect\ref{fig4}, \protect\ref{fig5},
and \protect\ref{fig7} for a more than half filled drum and
the $z^\prime=\protect\sqrt{1-z^2}$ transformation of the wedges.
$\theta=0.4\pi$. (See Eq.~(\protect\ref{M17})).
\label{fig8}}
\end{figure}

In the same manner can be considered case of the
$z^\prime=\sqrt{1-z^2}$ transformation for a more then half filled drum.
One can obtain easily

\begin{eqnarray}
\label{M17}
t_{2n}=4n\pi-2n\arctan(z\tan\theta)-2n\arctan(\sqrt{1-z^2}\tan\theta)
\phantom{eeeeeeeeeeeee}\nonumber\\
t_{2n+1}=(2n+1)2\pi-
2n\arctan(z\tan\theta)-(2n+1)\arctan(\sqrt{1-z^2}\tan\theta)
\end{eqnarray}
(see Fig.~\ref{fig8}).
The averaged concentration of the black fraction will be

\begin{equation}
\label{M18}
c(z)=\frac{\gamma}
{2\pi-\arctan(z\tan\theta)-\arctan(\sqrt{1-z^2}\tan\theta)}
\end{equation}
(cf Eq.~(\ref{M9})). Note that in the event of the
$z^\prime=\sqrt{1-z^2}$ transformation, mixing is far poorer then for
the $z^\prime=z$ transformation.

In conclusion, using discrete mapping approach,
we considered analytically two limit regimes of
avalanche mixing, when the mixing in the wedges is absolutely absent.
Nevertheless, finally, the material appears to be rather effectively
mixed. The experimental situation \cite{metcal} -- the mixing of salt
grains -- is between full mixing in wedges and what we call the
$z^\prime=z$ transformation of the wedges (see the experimental patterns of
the mixing in the paper of G.~Metcalfe {\it et al}). There was a question
\cite{dor1,dor2}: why is the experiment so close
to the extremal (the fastest!) regime of avalanche mixing, in which
granules fully mix in avalanches?  Why do the theory for this extremal
regime give such a surprising coincidence with the mixing time values
observing in the experiment? Now we can see the reason. As we have shown
in the present communication, the dependence of the quantity characterizing
the time of mixing on the drum filling
(see Fig.~\ref{fig6}) in the case of the
$z^\prime=z$ transformation appears to be rather similar to the behavior
of mixing time in the fastest possible for avalanche mixing regime.
Thus the experiment, which is between this two limit cases with a similar
behavior, can be described well.

I wish to thank V. V. Bryksin, Yu. A. Firsov,
A.~V.~Goltsev,  S.~A.~Ktitorov, E.~K.~Kudinov, B.~N.~Shalaev
for helpful discussions and B.~A.~Peratt and J.~A.~Yorke for sending
their paper \cite{peratt2} before the publication.
This work was supported in part by the RFBR grant.

\end{document}